\documentclass[pre,floatfix,twocolumn,showpacs,a4]{revtex4}

\usepackage[latin1]{inputenc}
\usepackage[T1]{fontenc}
\usepackage{ae,aecompl} 

\usepackage{graphicx}%
\usepackage{dcolumn}
\usepackage{tabularx}
\usepackage{epsfig}
\usepackage{subfigure}
\usepackage{amsmath, amsthm, amssymb}
\begin{document}

\newcommand{\be}{\begin{equation}}
\newcommand{\ee}{\end{equation}}
\newcommand{\bea}{\begin{eqnarray}}
\newcommand{\eea}{\end{eqnarray}}
\newcommand{\link}{\ensuremath{l}}
\newcommand{\currcom}{\ensuremath{C_c} }
\newcommand{\origcom}{\ensuremath{C_o} }

\newtheorem{thm}{Theorem}
\newtheorem{cor}[thm]{Corollary}
\newtheorem{lem}[thm]{Lemma}
\newenvironment{definition}[1][Definition]{\begin{trivlist}
\item[\hskip \labelsep {\bfseries #1}]}{\end{trivlist}}

\title[Short Title]{A sequential algorithm for fast clique percolation}
\author{Jussi M. Kumpula$^1$}
\email{jkumpula@lce.hut.fi}
\author{Mikko Kivel{\"a}$^1$}
\author{Kimmo Kaski$^1$}
\author{Jari Saram\"{a}ki$^1$}
\affiliation{%
$^1$Department of Biomedical Engineering and Computational Science, Helsinki University of Technology, P.O. Box 9203, FIN-02015 HUT, Finland \\
}%

\date{\today}
\begin{abstract}

In complex network research clique percolation, introduced by Palla \emph{et al.}, is a
deterministic community detection method, which allows for overlapping
communities and is purely based on local topological properties of a
network. Here we present a sequential clique percolation algorithm
(SCP) to do  fast community detection in weighted and unweighted
networks, for cliques of a chosen size. This method is based on sequentially
inserting the constituent links to the network  and simultaneously keeping track of
the emerging community structure. Unlike existing algorithms, the SCP
method allows for detecting $k$-clique communities at multiple weight
thresholds in a single run, and can simultaneously produce a dendrogram
representation of hierarchical community structure. In sparse weighted
networks, the SCP algorithm can also be used for implementing the weighted clique percolation
method recently introduced by Farkas \emph{et al}. The computational time of the SCP
algorithm scales linearly with the number of $k$-cliques in the
network. As an example, the method is applied to a product association
network, revealing its nested community structure.

\end{abstract}

\maketitle

\section{Introduction}

Over the last decade, complex networks have become a standard framework in the study of complex systems~\cite{RefWorks:179,RefWorks:133}. 
The simplicity of the network representation, where the interactions and interacting elements are mapped
to links and nodes, respectively, facilitates its use on a number of systems, ranging from human societies 
to biological systems. One prominent feature of complex networks is related to their mesoscopic properties. 
Networks often display \emph{modular} structure,
\emph{i.e.}, are structured in terms of modules or communities, which are, in general, sets of densely interconnected nodes.
Such communities are often closely related to functional units of the system, 
for example  groups of individuals interacting with each other in society 
\cite{RefWorks:48,RefWorks:144,RefWorks:75,RefWorks:114}, or
functional modules in metabolic networks \cite{RefWorks:146,RefWorks:67,RefWorks:52}. 

The problem of detecting communities in complex networks has received a lot of attention
during the last years. This problem is twofold: first, there is no unique way to rigorously define what constitutes a community. For any definition, several choices have to be made: whether communities are defined using local or global network properties, whether nodes can participate
in several communities, and whether the definition allows for weighted networks and nested hierarchy of communities. Second,
any definition is useful in practice only if it can be reformulated as an algorithm which scales well enough to allow processing networks
of large enough size. As a result, a large number of community definitions and their algorithmic implementations have been proposed over the recent years~\cite{RefWorks:46,RefWorks:70,RefWorks:172,RefWorks:141,RefWorks:143,RefWorks:142}; for a review see~\cite{RefWorks:139}.

In this paper we focus on a fast algorithmic implementation of the \emph{clique percolation} (CP) method, originally introduced by Palla \emph{et al.}~\cite{RefWorks:52}. The CP method is deterministic and it is based solely on local topological properties, defining a $k$-clique community as a set of nodes belonging to adjacent $k$-cliques. This allows for overlapping communities, \emph{i.e.}, nodes having
multiple community memberships. The CP method has earlier been successfully applied to various community detection problems:
detection of protein communities related to cancer metastasis~\cite{RefWorks:200}, analysis of communities in co-authorship, 
word-association and protein-interaction networks \cite{RefWorks:52}, 
and time evolution of social groups \cite{RefWorks:114}. In contrary
to existing implementations \cite{RefWorks:257}, which detect $k$-clique communities for all values of $k$ by first finding
the maximal cliques by an exponentially scaling algorithm~\cite{RefWorks:52},
we focus on rapid detection of communities for a chosen value of $k$.
Our sequential clique percolation (SCP) algorithm is based on sequentially inserting links to the network and keeping track of the emerging community structure. It has specifically been designed for weighted networks containing hierarchical communities which are reflected in the link weights. When links are inserted in decreasing order of weight, the algorithm allows for detecting $k$-clique communities at chosen threshold levels in a single run and simultaneously produces a dendrogram representation of hierarchical community structure. In addition, the algorithm can be used for very fast community detection for unweighted networks. 

This paper is structured as follows: first, we present our algorithm for the simplest, unweighted case, and discuss its scaling properties. We then move on to detecting nested communities in weighted networks, applying the algorithm to a product association network generated from data on sellers and products on an online auction site. Finally, we discuss a variation of the algorithm which is based on ordering $k$-cliques according to their weighted properties, and present our conclusions.

\section{The SCP algorithm}

Let us begin by defining $k$-cliques and $k$-clique communities~\cite{RefWorks:52,RefWorks:78}:\\
\\
$\bullet$ A $k$-clique is a set of $k$ nodes which are all connected to each other. A $k$-clique community, or $k$-community, 
is a set of nodes which can be reached by a series of overlapping $k$-cliques, where overlap means that  the $k$-cliques
share $k-1$ nodes. \\

It should be noted that 2-cliques correspond to pairs of nodes connected by single links and 1-cliques to single nodes.
Given a network $\Gamma$, the goal is then to find the $k$-communities defined as above. 
In our case, we restrict ourselves to some specific values of $k$. 
Usually choosing $k=3$ or $k=4$ yields useful information, and currently
these values of $k$ have yielded, to our knowledge, the most relevant communities in practical applications \cite{RefWorks:52,RefWorks:114,RefWorks:173,RefWorks:200}.
Our algorithm is based on detecting and storing $k$-communities as they emerge and consolidate when links are sequentially
inserted into the network.  One can think of the process as first "removing" each link $\link$ from the network $\Gamma$,
and then inserting them back one by one. For unweighted networks, the links can be inserted in any order, 
whereas for weighted networks, it may be desirable to sort the links by weight.

Our algorithm for detecting $k$-communities consists of two phases: 
the first phase of the algorithm detects $k$-cliques which form when a link
is inserted. These are then fed to the second phase, which
keeps track of formation and merging of $k$-communities by processing the found $k$-cliques.
The two parts of the algorithm are described in detail below.

\subsection{Phase I: Detecting the $k$-cliques}

\begin{figure}[t!]
\begin{center}
\includegraphics[width=0.2\textwidth]{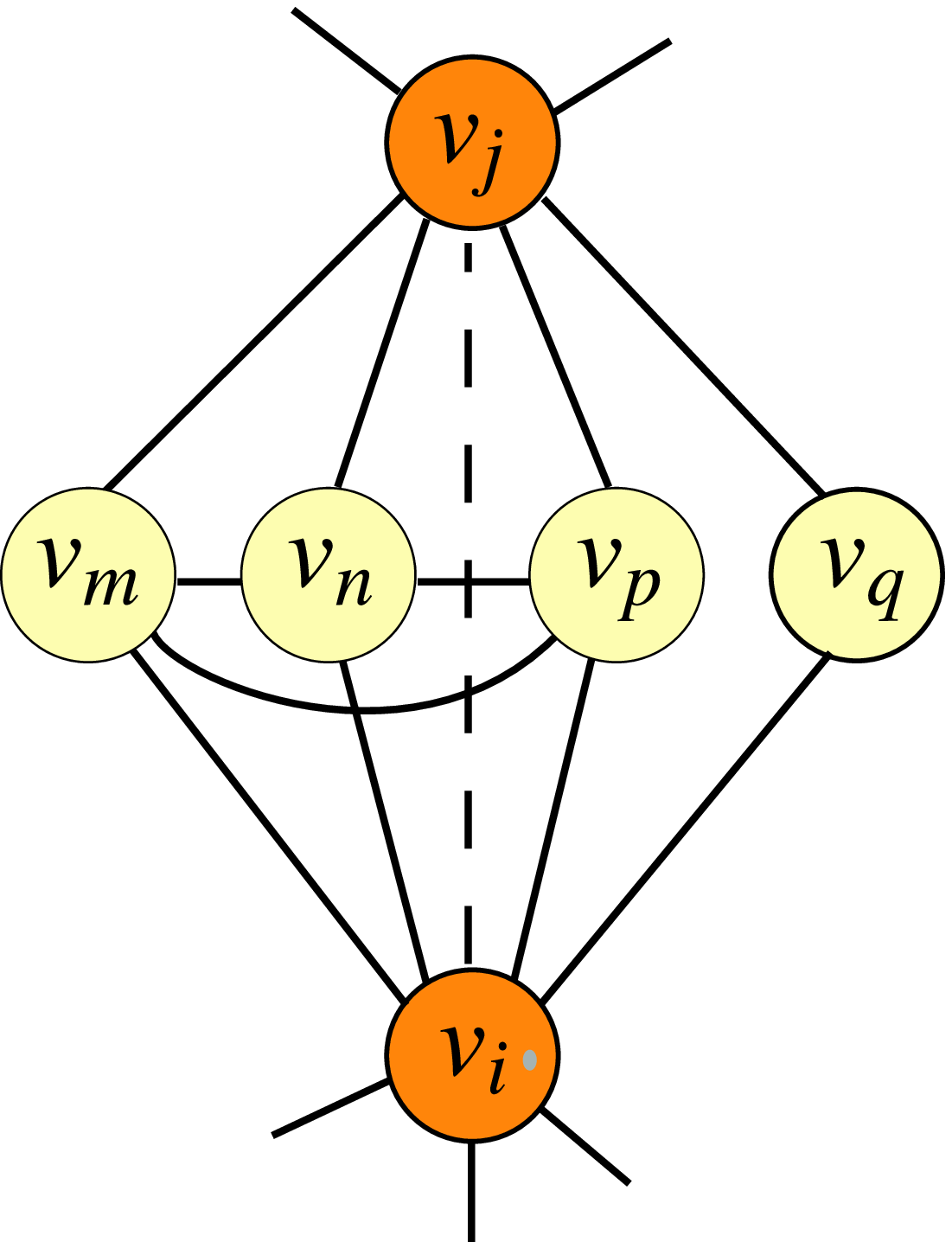} 
\caption{Schematic illustration of the process for detecting the $k$-cliques a newly inserted link completes. 
The dashed line depicts the new link, inserted between nodes $v_i$ and $v_j$.
The common neighbors of nodes $v_i$ and $v_j$ are $\mathcal{N}_{ij}=\left\{v_m,v_n,v_p,v_q\right\}$. 
For detecting newly formed 4-cliques, all pairs of nodes in $\mathcal{N}_{ij}$ are checked
to see if they are connected, that is, if they form a 2-clique. 
Each 2-clique in the set gives rise to a 4-clique, so 
in total the link $\link_{ij}$ will generate three 4-cliques.
In the case $k=5$, only one $3$-clique is found, which contains the nodes $v_m$, $v_n$ and $v_p$. 
It will give rise to a single $5$-clique including these nodes in addition to $v_i$ and $v_j$.
}
\label{fig:newClique}
\end{center}
\end{figure}

The first part of the algorithm involves detecting $k$-cliques which are formed when a link is inserted into the network.
Suppose now that the inserted link connects nodes $v_i$ and $v_j$ (see Fig.~\ref{fig:newClique}). 
The minimum requirement for a new $k$-clique to form is that nodes $v_i$ and $v_j$ both have degree of at least $k-1$.
 If this is the case, the algorithm proceeds by
collecting all nodes that are neighbors of both nodes, $\mathcal{N}_{ij}=\mathcal{N}_i \cap \mathcal{N}_j$, where $\mathcal{N}$ denotes neighborhood.  
Now, when the link $\link_{ij}$ is added, each $(k-2)$-clique contained in the set $\mathcal{N}_{ij}$ will give rise to a new $k$-clique.
Therefore, all newly formed $k$-cliques are found by detecting all the $(k-2)$-cliques in the $\mathcal{N}_{ij}$. For commonly used 
small clique sizes, this is very fast: for 3-cliques, 
$(k-2)$-cliques are single nodes, while for $k=4$, all connected
pairs of nodes in $\mathcal{N}_{ij}$ give rise to a new 4-clique.

Next the $k$-cliques detected as above are fed one by one into the second phase of the algorithm.

\begin{figure}[!tb]
\begin{center}
\includegraphics[width=0.5\textwidth]{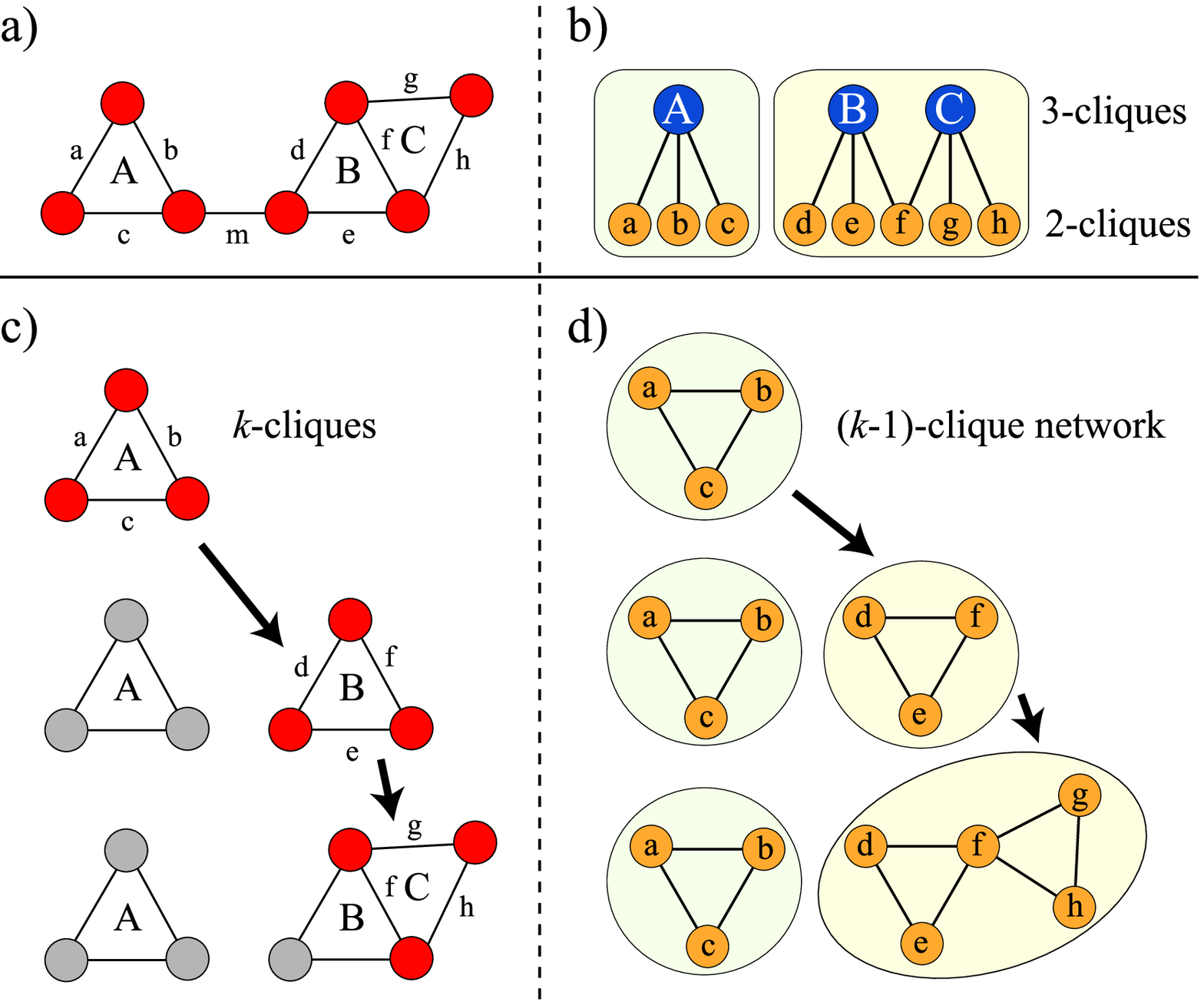} 
\caption{Illustration of the algorithm for detecting $k$-clique communities in a simple 
example network. Here, $k=3$. a) The original network $\Gamma$ consists of three 3-cliques labeled $A$, $B$, and $C$.
2-cliques, \emph{i.e.}, nodes connected by single links, are labeled with lower case letters. 
b) Bipartite network presentation of the clique structure. Note
 that in the bipartite network, the 3-cliques $B$ and $C$, which form a 3-clique community, are connected by the shared
2-clique  $f$. Clique $A$ forms another $3$-clique community.
c) 3-cliques detected by the first part of the algorithm as links are sequentially inserted into the network. 
Each new $k$-clique is denoted by red nodes whereas nodes associated with existing $k$-cliques
appear gray. 
d) Corresponding updates to the  $(k-1)$-clique network $\Gamma^*$ as a result of the second part of the algorithm.
$k$-clique communities correspond to connected components of this network (shaded areas).}
\label{fig:projection}
\end{center}
\end{figure}

\subsection{Phase II: Detecting the $k$-communities}

The second phase of the algorithm detects and keeps track of $k$-communities which form and merge
when new $k$-cliques are input from the first phase. Because 
a $k$-community is defined as a set of nodes 
which all can be reached by
a series of overlapping $k$-cliques, the crucial issue here is the efficient detection of overlap between
$k$-cliques. 
A naive approach would be to search for shared sets of $k-1$ nodes between the newly input clique and 
all existing cliques. However, the required computational effort makes this approach unpractical. 
Instead, we take advantage of the sequential nature of the process by "locally" detecting possible 
overlap of each new $k$-clique with existing $k$-communities and by updating the community structure accordingly.
 
Let us begin by noting that the 
$k$-community structure of a network can be represented by 
a bipartite network, where the two types of nodes represent $k$-cliques and $(k-1)$-cliques. In this network,
a link exists between two nodes of different type if the $k$-clique contains the
$(k-1)$-clique as a sub-clique.
This is illustrated in Fig.~\ref{fig:projection}. 
The usefulness of this representation becomes apparent in the following: 
each connected component in this bipartite network
corresponds to a $k$-clique community, because by definition $k$-cliques belonging to the
same community are connected through shared $(k-1)$-cliques.
Furthermore, connected components of the unipartite projections of the bipartite network \footnote{In a unipartite projection, the bipartite network is collapsed such that only nodes of one type are left, each pair connected by a link if they are both connected to the same node(s) of the other type in the original bipartite network.} similarly
correspond to $k$-clique communities. In the following, we focus 
on the $(k-1$)-clique projection of this bipartite network. We denote the 
network resulting from this projection by $\Gamma^\ast$. In this unipartite network,
nodes $v^\ast$ represent the $(k-1)$-cliques of $\Gamma$, and links $\link^\ast$ exist between
nodes which are sub-cliques of the same $k$-clique.

For the sake of clarity, we will first present a "physical" interpretation of Phase II of the algorithm, and then discuss the algorithmic implementation where certain shortcuts can be made.
Similarly to Phase I, where the original network $\Gamma$ is reconstructed link by link,
Phase II of the SCP algorithm sequentially builds up $\Gamma^\ast$ from the $k$-cliques brought forward from Phase I. At the same time, it keeps track of the connected components of  $\Gamma^\ast$
(see Fig.~\ref{fig:projection}, panels c and d). These correspond to $k$-clique communities. 
When a new $k$-clique is input from Phase I, its constituent $(k-1)$-cliques are first extracted; 
obviously there are always $k$ of such sub-cliques. Each of these $(k-1)$ cliques corresponds
to a node in $\Gamma^\ast$. 
Some of these nodes may already be present, if the corresponding $(k-1)$-cliques have been handled earlier as part of another $k$-clique; if not, they are created at this stage. Finally, links
are created between members of this set of $k$ nodes, and resulting changes in the connected component structure
of $\Gamma^\ast$ are recorded.

In the algorithmic implementation, things can be done somewhat more efficiently, resembling
techniques used in link percolation. The actual network $\Gamma^\ast$ does
not need to be constructed, as it is enough to keep track of its connected components, \emph{i.e.},
the component indices of its nodes $v^\ast$. This
is equal to link percolation in $\Gamma^\ast$, which can be
implemented for example with disjoint-set forests \cite{RefWorks:256}. 
At this stage it is enough to ensure that all $(k-1)$-clique-nodes corresponding to the new $k$ clique are marked to
belong to the same component (the new $(k-1)$-cliques 
and their links may either form a new connected component, 
merge with an existing component, or join together at most $k$ existing components).

The above process is then repeated for each $k$-clique input from Phase I. Finally, 
once all links have been inserted (Phase I) and the subsequently formed $k$-cliques
handled (Phase II), the $k$-communities of the original network $\Gamma$ can be read from
the component indices of $v^\ast$, assigning 
nodes of $\Gamma$ to their corresponding communities.

In theory, it would also be possible to keep track of the connected components of the whole bipartite
network or alternatively project the bipartite network to $k$-cliques instead of $(k-1)$-cliques.
Both representations contain the same connected components and would thus yield the same $k$-clique communities.
However, 
the former alternative is unnecessarily complicated as it involves nodes of two types. The latter
implementation is not as computationally effective as the current choice in cases where a newly inserted $k$-clique
overlaps with a large number of existing $k$-cliques.

\subsection{Scaling of the algorithm}
\label{subsec:scaling}

\begin{figure*}[!htbp]
\begin{center}
\includegraphics[width=0.8\linewidth]{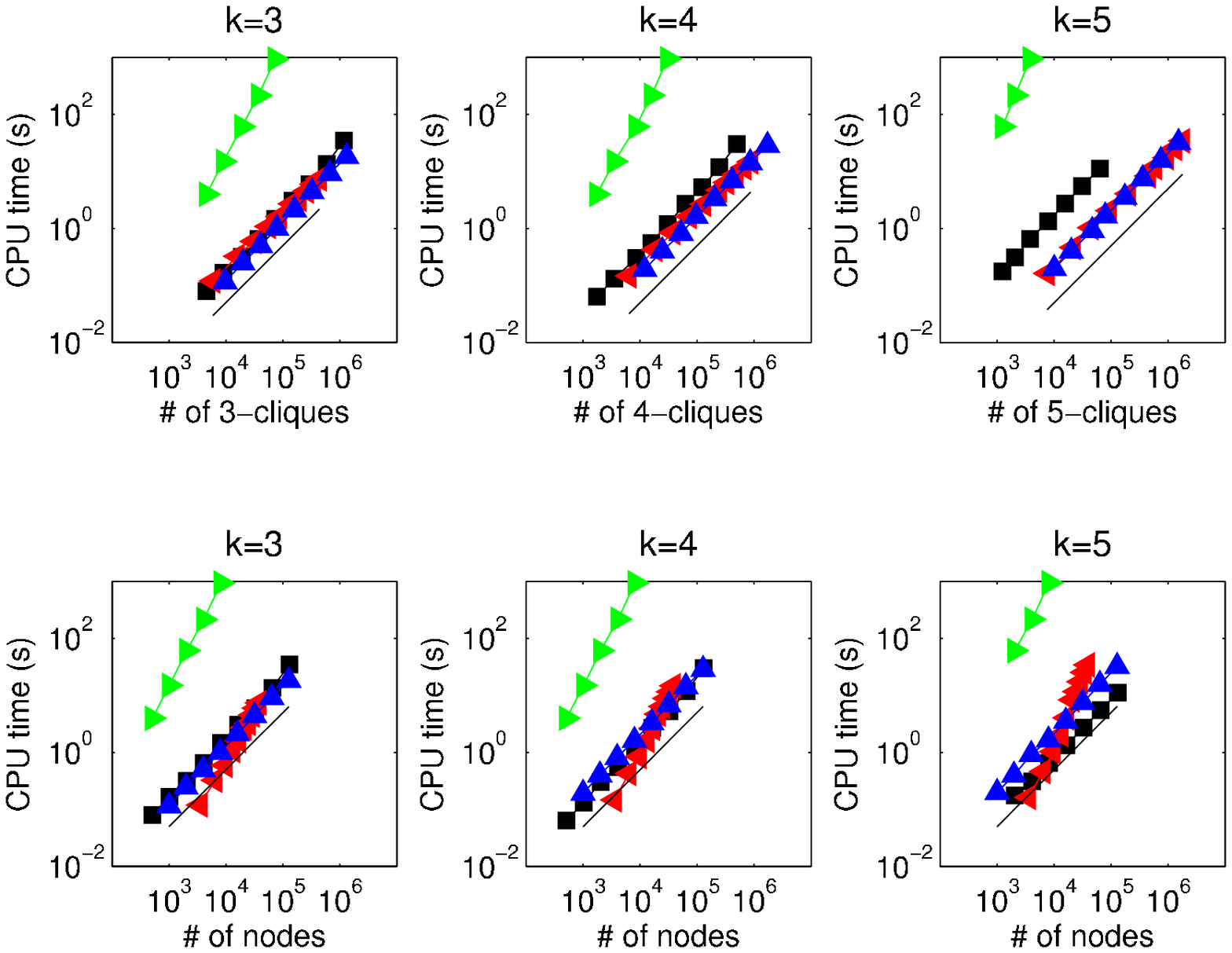} 
\caption{Computation time of the algorithm for three values of $k$, as a function of the number of $k$-cliques (upper row) and network size (lower row). Symbols denote different test networks: GN ($\blacksquare$), WSN ($\blacktriangle$), and
CM ($\blacktriangleleft$), see text for details.
The solid line is a linear reference. For comparison, we have also plotted the computational time of the CFinder 1.21 algorithm  for the GN networks 
($\blacktriangleright$).
Note that CFinder always processes all values of $k$. }
\label{fig:cpu_time}
\end{center}
\end{figure*}

Let us next discuss the performance of the SCP algorithm, before moving on to its application to weighted network analysis. Obviously, the computational time required to process a network depends on its properties; here, we wish to investigate the performance as a function of network size and the number of $k$-cliques contained in the network. To do this, we have applied the SCP algorithm on three types of networks with adjustable sizes. The first test case (GN), introduced by Girvan and Newman~\cite{RefWorks:48}, contains built-in communities and has often been used for similar purposes. The GN networks used here consist
of groups of 32 nodes, where each node has on the average 12 links to nodes of the same group and 4 links to other groups. The network size $N$ is varied by changing the number of such groups. The second type  of networks (WSN) 
is generated using a recently published model of weighted social networks with communities~\cite{RefWorks:112}, 
using parameter values similar to the original reference. As the third type, we have used co-authorship networks based on the \emph{cond-mat} archive (CM), constructed similarly to e.g.~\cite{RefWorks:167}. However, in order to vary the network size, we have used time windows of varying length, such that two authors are connected if they have published a joint paper during the time window. It should be noted that although the WSN networks are inherently weighted, and the CM networks can also be considered such, here we consider binary versions of both types for the performance analysis.

Results in Fig.~\ref{fig:cpu_time} show that the computational time of the SCP algorithm
grows practically linearly
as a function of the number of $k$-cliques for all networks. This is as expected, because the computational time of the
algorithm is dominated by the process of detecting $k$-cliques and processing them for overlap, such that
each $k$-clique is processed exactly twice.
This is also reflected in the network size dependence of the required computational time for both types of model networks
(GN, WSN). For these networks  the local structure remains essentially unchanged as the network grows and it appears that the
number of $k$-cliques grows linearly with $N$. 
However, for the CM networks, the computational time grows faster than linearly as a function of 
network size. This is because the CM network is a projection of a bipartite author-publication network 
containing large cliques that grow in size when N increases. 
The problem is, as pointed out by Palla \emph{et al.}~\cite{RefWorks:52}, 
 that the number of sub-cliques of size $k$ within a clique of size $s$ is $\binom{s}{k}$.
In the limit $s \gg k$ this leads to
\be
\binom{s}{k} \approx \frac {s^k}{k!}.
\ee
Hence for large $s$, the number of $k$-cliques grows
as $k^{th}$ power of $s$, meaning that for networks containing large cliques 
the SCP method performs best for rather small values of $k$. 
For example, when $k>10$ the analysis of the largest CM networks becomes extremely slow with the SCP method. However, when very large cliques are not abundant in the network under investigation, the SCP algorithm is very fast
even for networks of large size. For example, detecting 4-clique communities
in a mobile phone call network having approximately 4 million nodes and 6 million links \cite{RefWorks:129}
takes approximately one minute on a standard desktop computer. 
Thus, for networks where cliques are on the average fairly small, the main practical limitations of this algorithm seem to be related to the memory
consumption as it requires keeping all $(k-1)$-cliques of the network in memory.  

Finally, let us compare the performance of the SCP algorithm and the existing method 
(CFinder 1.21, \cite{RefWorks:52}). Evidently, this comparison is somewhat complicated,
as CFinder simultaneously processes all clique sizes, whereas the SCP algorithm is by construction
limited to a single value of $k$. Nevertheless, summing
up the processing times for all values of $k$, we have observed that for the GN network,
the processing time of the SCP algorithm scales linearly with network size, whereas
CFinder 1.21 appears to scale as $N^2$ (see Fig.~\ref{fig:cpu_time}). However, for denser networks, such
as the CM network, the comparison becomes somewhat meaningless as both methods
become extraordinarily slow. This is due to the very large number of $k$-cliques
as discussed above.
It should be noted here that the unpublished beta version, CFinder 2.0b, appears
to scale far better than CFinder 1.21 and seems to be able to deal with very large cliques.
However, the key strength of the SCP algorithm is its speed in weighted network analysis: it is
able to process multiple weight thresholds in a single run (see Section \ref{multith} below). 
With the earlier method, this quickly becomes unfeasible, as the networks corresponding to each threshold have to be separately input and analyzed. Thus, even if the processing time of 
both methods would be exactly the same for a single network, obtaining the $k$-community structure
for 100 weight thresholds would be 100 times faster with the SCP algorithm. Another important
difference is the inherent ability of the SCP method to produce a dendrogram of nested $k$-communities; this feature does not exist in earlier implementations (again, see Section \ref{multith} below).

\section{SCP for weighted networks}
\subsection{Thresholding and nested communities}
\label{multith}

Let us move on to weighted networks, where the concept of community structure becomes somewhat
more complicated. Perhaps only for the very simplest cases, where the networks are sparse, weights can
be disregarded, such that communities are associated with the pure topology of the network. However, this is 
usually not feasible, as weighted networks can be rather dense, even to such an extent
that the topology no longer matters, as any modular structure is encoded in the link weights only.
This is the case for example in stock interaction networks~\cite{RefWorks:175}, whose natural representation is a weight matrix
with only nonzero elements. 

For such networks, one is essentially left with two choices: the first is to threshold the network, 
such that links whose weights are considered insignificantly small are removed and communities in the resulting sparse network are detected. It is evident 
that choosing the right threshold is a non-trivial task; in fact,  for many cases it may be
better to take a multi-resolution approach, by investigating the resulting community structure for a range
of thresholds. Another option is to consider the weights directly when defining what constitutes a community, and 
to apply a method which is based on this definition \cite{RefWorks:173,RefWorks:175}.

\begin{figure*}[htbp]
  \begin{center}

\includegraphics[width=0.9\textwidth]{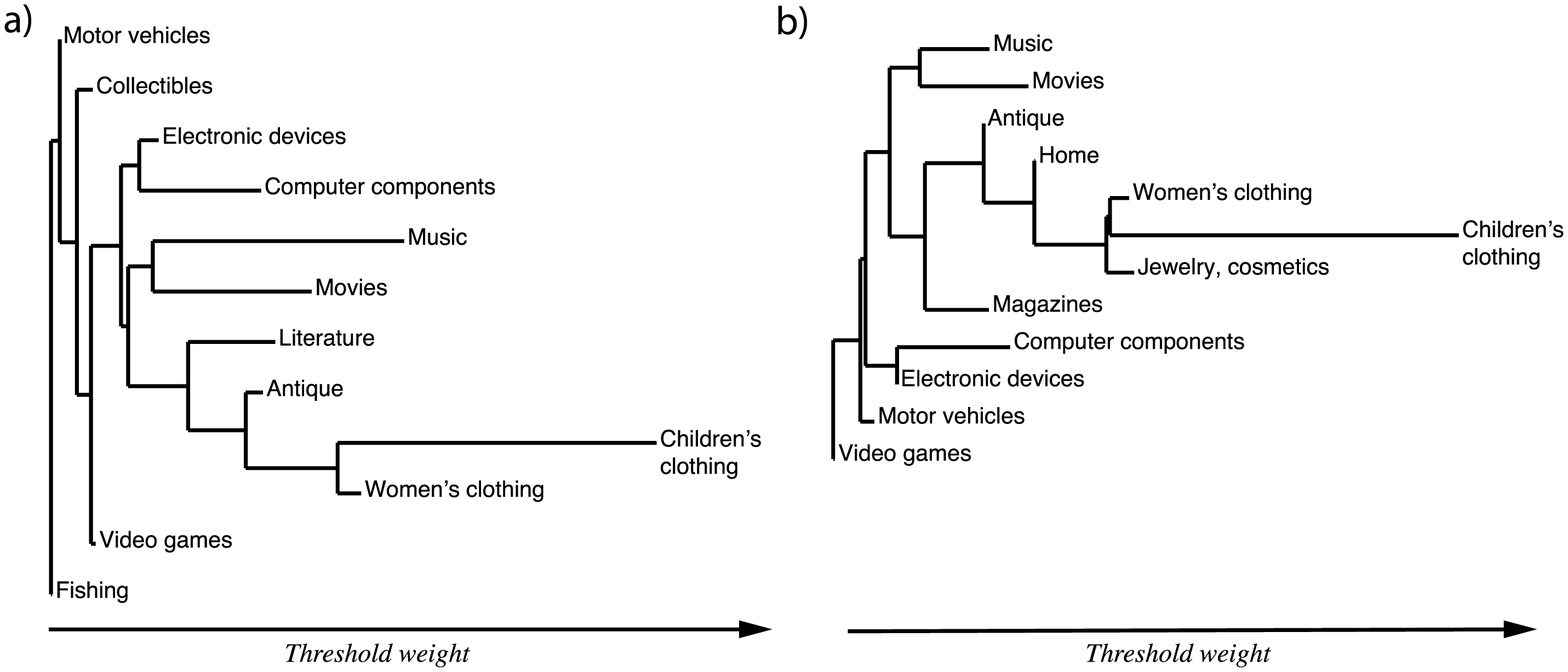}
    \caption{Dendrogram visualization of the nested $k$-community structure of the trading categories of the Finnish online auction site Huuto.net for $k=3$ (a) and $k=4$ (b).}
    \label{fig:dendrogram}
  \end{center}
\end{figure*}

In the original formulation of the clique percolation algorithm, Palla {\it et al.} suggested a rule for choosing a weight 
threshold $w^*$  for the network, such that  the resulting $k$-community structure would be as diverse as 
possible \cite{RefWorks:52}. More specifically, $w^*$ is chosen such that the largest community is twice the size of the second largest
one, \emph{i.e.}, below the percolation threshold where a giant $k$-clique community appears. For the original implementation, 
the algorithm had to be run from the beginning for each threshold level. One of the benefits of our approach is that it allows for obtaining the $k$-communities at any point of the process of adding links, which is just thresholding done in reverse: 
If the links of the original network $\Gamma$ are sorted and processed in descending order of weight, the algorithm
yields for each link the $k$-community structure of $\Gamma$ thresholded by the weight of the link.
This is very useful for selecting the threshold, as all threshold values can be processed in a single run.
Note that for dense networks, sweeping through the entire range of weights is not needed:
the algorithm can be stopped before (or immediately after) communities are entirely "smeared
out" by a giant community.
Stopping the algorithm in time can greatly reduce the workload in
dense networks as usually only a small fraction of all links need to be added
before the percolating component is found, after which adding more
links does not increase the number of nodes in the
communities, but only makes the community denser in cliques.

However, by focusing on a single threshold weight, valuable information of the community structure contained
in the correlations between weights can be lost. Often, the modular structure of networks is inherently \emph{hierarchical} -- 
denser and stronger communities are nested inside weaker ones, which may further be embedded inside even weaker 
ones~\cite{RefWorks:253,RefWorks:174,RefWorks:142,RefWorks:254}.
It is then natural to investigate this nestedness by considering the development of the community structure when
the weight threshold is swept through the range of interest. Evidently, this requires book-keeping of the emergence
and merging of communities as the threshold is progressively lowered. For the SCP algorithm, this book-keeping is 
inbuilt: all necessary information can directly be recorded in Phase II of the algorithm. In particular, it is easy to
store  when a $k$-community appears, which nodes belong to it, how its size grows as new
$k$-cliques join it, and when it merges with other $k$-communities. It should be stressed here that
this is a genuine advantage: separately detecting the community structure for each threshold and then tracking the formation and merging of communities would be very difficult and time-consuming.

This information on the nested community 
structure is best visualized with a 
\emph{dendrogram}, which is a common presentation format in agglomerative community detection (see, e.g., \cite{RefWorks:254}). 
In a dendrogram, horizontal lines correspond to communities, and a branching of the lines denotes
communities merging.  Choosing a single weight threshold would correspond to taking a vertical slice
of the dendrogram.  Fig.~\ref{fig:dendrogram} shows two examples of the nested community structure
within a product category network, for $k=3$ and $k=4$. This network is constructed from online
trading data, downloaded from the  Finnish auction website Huuto.net. In this network, 
nodes correspond to product categories ($N=345$), and the weights of links connecting two categories 
to the number of individuals who have been trading in both of them. This network is very dense,
the number of links is 52536, corresponding to a link density $\rho = 0.89$, and thus the network can be considered as 
a suitable test case for the evolution of community structure while sweeping the threshold weight.
In Fig.~\ref{fig:dendrogram} the labels associated with each community describe their dominant product categories.
Although the dendrograms formed by using $k=3$ and $k=4$ are not identical,
several similar communities appear for both values. From the commonsensical point of view,
these appear natural: electronic devices and computer components merge to 
a single community, as do music and movies, and children's  and women's clothing. 

Often it is not possible nor meaningful to include all $k$-communities
in such a visualization: the outcome would be too complicated to be interpreted by visual inspection.
The main problem are the numerous single $k$-cliques, which merge to
larger $k$-communities. 
For any analysis of the dendrogram structure the entire data should be used but for
visualization purposes it is useful to threshold the dendrogram such that
only $k$-communities which are larger than a threshold size $N_{th}$ appear in the plot.
In Fig.~\ref{fig:dendrogram} $k$-communities of sizes larger than $k$ are displayed, i.e., 
$N_{th}=k$. 

\subsection{Weighted $k$-clique percolation}
\label{WCP}
As pointed out above, considering the weights in the definition of what constitutes a community
is an alternative to simply discarding low-weight links. Such an extension for clique percolation
has recently been introduced by Farkas \emph{et al.}~in \cite{RefWorks:173}. In this method, each $k$-clique
is assigned a "weight", which equals the \emph{intensity}~\cite{RefWorks:240} of its edge weights. 
The intensity is defined as the geometric mean of the link weights in the $k$-clique.
The community structure is then obtained by choosing an intensity threshold $I^*$
and taking into account only those $k$-cliques whose intensity is above $I^*$.

For our SCP algorithm, a simple modification allows for weighted clique percolation
according to the above scheme. To achieve this, instead of building the
$k$-communities simultaneously as the $k$-cliques emerge, all links are first inserted to the network and the
resulting $k$-cliques are stored. Then, the intensity of each of these $k$-cliques
is calculated, and the cliques are sorted with respect to the intensity.
Finally, the sorted $k$-cliques are processed one by one by the second part of the algorithm,
until the intensity threshold is reached. Multiple thresholding levels are obtained as before, 
but now with respect to $k$-clique intensities, and a dendrogram can be constructed similarly.
Note that in addition to intensity, any other measure describing the "weight" of the cliques
can be used, \emph{e.g.}, if homogeneous cliques are sought for, one could also take
the clique \emph{coherence}~\cite{RefWorks:240} into account. 
Sorting cliques according to their intensities was briefly described by Farkas et al. in \cite{RefWorks:173}; their construction appears somewhat similar to ours as the intensity-sorted cliques are handled in succession, and the method for obtaining overlapping $k$-communities seems to correspond to building the whole bipartite network between $k$- and $(k-1)$-cliques.

The above procedure requires keeping all $k$-cliques in the memory
in addition to the $(k-1)$-cliques.
In most cases the loss of speed is minimal, as the additional
computational load is related to the memory consumption and sorting of cliques,
which can be done in log-linear time. 
However, a possible problem related to the SCP algorithm -- and the weighted clique percolation method in general --
is that all $k$-cliques have to be processed individually,  and their number
can be very large in dense networks as discussed in Section \ref{subsec:scaling}. When the link weight thresholding procedure
of Section \ref{multith}
is applied, this problem can be somewhat circumvented by simply stopping the 
algorithm as soon as enough links have been inserted for obtaining the community
structure at the desired "resolution". However, for intensity-based clique percolation
this cannot be done, as all $k$-cliques have to be detected and sorted first.

\section{Conclusions}
We have introduced a sequential clique percolation algorithm for detecting $k$-clique communities
in a network by sequentially inserting its edges and keeping track of the 
emerging community structure \footnote{A Python implementation of the algorithm can be found online at {\it http://www.lce.hut.fi/$\sim$mtkivela/kclique.html}}. This algorithm has specifically been designed
for (dense) weighted networks, where weight-based thresholding of either the
links or the cliques formed by them is necessary for obtaining meaningful
information on the structure. By applying the algorithm on test networks,
we have shown that the computational time required to process a network 
scales linearly with the number of $k$-cliques in the network.
The sequential nature of the algorithm allows run-time construction of
a dendrogram presentation of the nested hierarchical $k$-community structure,
which we have illustrated using a product category network.
 
The main tradeoff for our algorithm is that it detects the $k$-communities
for a chosen value of $k$ with multiple weight thresholds in a single run, instead of 
obtaining $k$-communities for all values of $k$ with a single weight threshold as is done in the 
maximal clique algorithms.
Hence the SCP algorithm can be considered complementary
to earlier presented solutions~ \cite{RefWorks:52} . Neither of these algorithms can be argued to 
be strictly better or faster than the other as their performance depends heavily on the
network topology and other aspects of the problem they are solving. The SCP 
 algorithm is particularly useful when a
small clique size $k$ is used and when multiple
weight threshold levels need to be studied, or no prior knowledge of the proper threshold
level of a dense weighted network is at hand.
The algorithm can also be considered as a reasonable choice for very large
sparse networks as suggested by the short computation times of
the community structure of a mobile telephony network having millions of nodes and links.

\textbf{Acknowledgements:} We thank J.~Hyv\"onen and J.~Kert\'esz for useful discussions, and acknowledge programming assistance
by J.~Hyv\"onen. We acknowledge support by the Academy of Finland, the Finnish Center of Excellence program 2006-2011, project no.~213470.
J.K. is partly supported by the GETA graduate school. J.S. and M.K. acknowledge support by the European Commission NEST Pathfinder initiative on Complexity through project EDEN (Contract 043251).


\bibliography{kumpula_ref}

\end{document}